\documentclass[twocolumn,preprintnumbers,floatfix,superscriptaddress,nofootinbib]{revtex4}
\usepackage[utf8]{inputenc}
\usepackage{setspace}
\usepackage{booktabs}
\usepackage{rotating}
\usepackage{url,comment}
\usepackage{times}
\usepackage{latexsym}
\usepackage{graphicx, graphics, amssymb, slashed, xcolor,amsthm, array}
\usepackage{subfigure}
\usepackage{listings} 
\usepackage{afterpage}
\usepackage{physics}
\usepackage[colorlinks=true,linkcolor=blue,citecolor=blue,urlcolor=blue]{hyperref}
\usepackage{here}
\usepackage{MnSymbol} 
\usepackage{braket}

\newcommand\nn{\nonumber}
\newcommand\bea{\begin{equation} \begin{aligned}}
\newcommand\eea{\end{aligned} \end{equation}}

\newcommand\B{{\tilde{A}}}

\begin{document}

%\vspace*{-30mm}

\title{A Mass for the Dual Photon}

%%%%%%%%%%%%%%%%%%%%%%%%%%%%%%%%%%%%%%%%%%%%%%%%%%%%%%%

\author{Anson Hook}
\email{hook@umd.edu}
\affiliation{Maryland Center for Fundamental Physics, Department of Physics, University of Maryland, College Park, MD 20742, U.S.A.}
\affiliation{Perimeter Institute for Theoretical Physics, Waterloo, Ontario N2L 2Y5, Canada}

\author{Junwu Huang}
\email{jhuang@perimeterinstitute.ca}
\affiliation{Perimeter Institute for Theoretical Physics, Waterloo, Ontario N2L 2Y5, Canada}

%%%%%%%%%%%%%%%%%%%%%%%%%%%%%%%%%%%%%%%%%%%%%%%%%%%%%%%

\vspace*{1cm}

\begin{abstract}

We explore a novel IR phase of electromagnetism and place constraints on it.
The usual IR modification of electromagnetism, the Higgs phase, involves adding a photon mass for the gauge field $A_\mu$, which screens electric fields and confines magnetic fields.  We explore the confined phase resulting from adding a mass term for the dual photon, which screens magnetic fields and confines electric fields.
We study the theory of a dual photon mass and argue that it is a consistent effective field theory.  We then elucidate the phenomenological consequences of such a mass term and derive constraints on it.  As the current constraints come with large uncertainties, we also propose a few new searches for a dual photon mass term.

\end{abstract}

\maketitle

\section{Introduction}

Electromagnetism is a main pillar of physics and quantum electrodynamics (QED) is one of the best tested theories we know. The force carrier, the photon, famously is the only massless particle in the Standard Model.  Other particles, such as neutrinos, were originally thought to be massless but have since been demonstrated experimentally to be massive. This fact has led people to consider theories with a massive photon, described by the Proca/Stueckelberg Lagrangian~\cite{Stueckelberg:1938hvi,Ruegg:2003ps}, and to test experimentally if the photon is indeed massless, or if the mass is just too small to be measured.

Electromagnetic duality has been historically important.  It led to the discovery of Faraday's Law and was formalized in quantum theories with Montonen–Olive duality~\cite{Montonen:1977sn}.  
In QED, this symmetry is spontaneously broken by the fact that we have electrically charged particles but, at least so far, no magnetically charged particles. A Proca/Stueckelberg mass of the photon also spontaneously breaks this symmetry, as it screens  electric fields but confines magnetic fields.  However, there is no {\it a priori} reason for the two breakings to be {\it aligned} (mutually local). As such, we consider the electromagnetic dual of a photon mass, a dual photon mass.

In this article, we propose a new IR modification to QED, the inclusion of a mass term for the dual photon ($\B_\mu$) rather than the usual photon ($A_\mu$).  The dual photon is related to the usual photon by the standard electromagnetic duality relationship~\footnote{Following Ref.~\cite{Zwanziger:1970hk} we will make the following simplifications $(a \wedge b) = a^\mu b^\nu - b^\mu a^\nu$, $a \cdot F = a_\mu F^{\mu \nu}$ and $\overline F = \frac{1}{2} \epsilon^{\mu \nu \rho \sigma} F_{\rho \sigma}$.}
\bea \label{Eq: dual}
F_A = \partial \wedge A \qquad F_\B = \partial \wedge \B \qquad F_\B = \overline{F}_A .
\eea
Electrically (magnetically) charged particles couple to the gauge field $A_\mu$ ($\B_\mu$).  Our modification involves adding to the QED Lagrangian a term
\bea \label{Eq: modification}
\delta \mathcal{L}  = \frac{1}{2} \tilde m^2 \B_\mu \B^\mu .
\eea
As mentioned before, this new modification is only different from a Stueckelberg mass when the theory contains the electron.  In the absence of an electron, what is an electric field and what is a magnetic field is merely a matter of convention.  Thus, in the absence of sources, mass terms for the gauge field and the dual gauge field are completely equivalent.  However, once the electron has been included into the theory and has been defined to carry electric charge, the two different mass terms are physically inequivalent.

Whenever one discusses electrically and magnetically charged particles at the same time, there have been historical concerns as well as historical resolutions.  The first concern is that the relationship between $\B_\mu$ and $A_\mu$ is not local and that including both at the same time would result in a non-local theory.  If one is willing to give up explicit Lorentz invariance, one can write down a local Lagrangian describing a theory of electrically and magnetically charged particles~\cite{Zwanziger:1970hk}.  The lack of explicit Lorentz invariance is the direction of the infamous Dirac string~\cite{Dirac:1948um}.  Requiring that the Dirac string is unphysical, which restores Lorentz invariance, leads to the quantization of electric and magnetic charges $\frac{e g}{2 \pi} = \mathbb{Z}$.

The second concern is a direct result of charge quantization.  If a theory has electric and magnetic charges, one of the two theories is necessarily strongly coupled.  It is difficult to prove anything about a strongly coupled theory and thus we are led to consider whether the modification Eq.~\ref{Eq: modification} is inherently strongly coupled.  Luckily for us, while we have involved the dual photon, we have not added any magnetically charged particles, so the magnetic coupling $g$ does not make an appearance.  As we will see below, the theory is still weakly coupled and completely calculable.

While we are considering the unstudied case of a Stueckelberg mass for the dual photon, there are quite a number of results on a mass term that is the result of the dual of a Higgs mechanism.  The first example is a symmetry breaking structure $SU(2)/U(1)/\varnothing$.  The first symmetry breaking comes from an adjoint scalar and produces 't Hooft-Polyakov monopoles~\cite{tHooft:1974kcl,Polyakov:1974ek}.  The second symmetry breaking comes from the standard Higgs mechanism generating a mass for the photon.  Switching the definition of electric and magnetic via electromagnetic duality, we have a theory with an electron and a magnetically Higgsed gauge boson.  

With supersymmetry, even more powerful results can be proven.  Seiberg-Witten demonstrated in particular cases that confinement of non-abelian gauge theories is the result of a magnetic Higgs mechanism~\cite{Seiberg:1994rs}.  In specific Seiberg-Witten examples called Argyres-Douglas models, it was shown that the IR theory contains massless electrically and magnetically charged particles~\cite{Argyres:1995jj}.  The strongly coupled theory was shown to be a conformal field theory and deformation of the UV theory was argued to lead to a magnetic Higgs mechanism.

Despite the fact that we know theories with electric and magnetic charges exist, our ability to write a Lagrangian for these theories is extremely limited.  The best attempt that we know of is Zwanziger's two potential formalism which is a first quantized theory~\cite{Zwanziger:1970hk}.  Particle creation and annihilation cannot be described using that formalism.  If one attempts to second quantize these theories, the end result is a Lorentz violating theory.  However, this limitation is well-known and does not invalidate the theory~\cite{Joe}.

As we do not wish to include magnetically charged particles, only a dual photon mass, things are marginally better~\footnote{At long distances, the theory is just a theory of open strings with electrons and protons at the end points.  The bounds on the dual photon mass will be strong enough that this simple theory will not be sufficient for what follows.}.  While our Lagrangian is still Lorentz violating, the Lorentz violation has a simple physical understanding.  In these theories, the Lorentz violation is the direction of the unphysical Dirac string.  However, when the dual photon mass is non-zero, the Dirac string develops an electromagnetic dressing which is the physical observable flux tube.  
Because of the dual photon mass, the theory attempts to screen the flux carried by the Dirac string, which is the origin of the physical string giving confinement.  The direction of the Dirac string specifies the direction of the string to which any given electron is connected.  Changing this direction corresponds to physically different scenarios, which is why the theory appears to be Lorentz violating.  
Generic vortex lines are dynamical and relax as the electrons move. In our Lagrangian, the vortex lines are non-dynamical and thus we can only describe steady states configurations~\cite{Ruegg:2003ps}.  The Lagrangians that we present in Sec.~\ref{Sec: L} will only be useful in a steady state first quantized picture when all of the straight Dirac strings point in the same direction. We leave studies with a dynamical fluid of vortex lines to future work.

In what follows, we first describe the Lagrangian that we will be utilizing in Sec.~\ref{Sec: L} and review some of the formalism used when describing electrically and magnetically charged particles.  In Sec.~\ref{Sec: pheno}, we describe some of the physical effects that result from a non-zero dual photon mass.  In Sec.~\ref{Sec: constraint} we place constraints on $\tilde m$ and propose a new experiment to search for it.  We discuss conclusions in Sec.~\ref{Sec: conclusion}.

\section{A Lagrangian for a confining $U(1)$ theory}  \label{Sec: L}

We will present three different but equivalent ways in which to introduce our mass term in the Lagrangian.  When using just the gauge field $A_\mu$, our mass term will be explicitly non-local and Lorentz symmetry breaking.  When using just the gauge field $\B_\mu$, our mass term will be local and Lorentz symmetry preserving, but the coupling of $\B_\mu$ to the electron will be non-local and Lorentz symmetry breaking.  When using both gauge fields $A_\mu$ and $\B_\mu$, our Lagrangian is local but the kinetic terms of $A_\mu$ and $\B_\mu$ will break Lorentz symmetry.  In all cases, the breaking of Lorentz invariance will be done with a space-like vector $n^\mu$ that corresponds to the direction of a straight Dirac string. Some of what we discuss is known in some shape or form in the condensed matter literature, and we found Ref.~\cite{Ripka:2003vv} to be a useful review of the subject.

\paragraph{A non-local mass term}

When written in terms of $A_\mu$, our Lagrangian is simply
\bea \label{Eq: one}
\mathcal{L} = -\frac{1}{4} &F^2 + A_\mu J^\mu \\
& - \frac{1}{2}  \left ( n \cdot \overline{\partial \wedge A}\right ) \frac{n^2 \,\tilde m^2}{n^2 \,\tilde m^2 + ( n \cdot \partial )^2} \left ( n \cdot \overline{\partial \wedge A}\right ) \, .
\eea 
This mass term is explicitly non-local due to the derivatives in the denominator, but the interaction between the photon and the standard electric current $J$ is local.  This Lagrangian highlights how the theory reverts back to standard QED when $\tilde m \rightarrow 0$~\footnote{Without a second quantized theory, we cannot specify how exactly a dual photon mass emerges from the full $SU(2)_W \times U(1)_Y$ invariant theory.  From degree of freedom counting, we expect that a dual photon mass term for $U(1)_Y$ will result in a dual photon mass term for electromagnetism at low energies.}.

\paragraph{A non-local interaction}

If we use the dual photon $\B_\mu$, the non-locality can be shifted from the mass term to the coupling with the electrons.  This formalism is the same as what Dirac used when describing monopoles.  In this basis, the Lagrangian is
\bea \label{Eq: two}
\mathcal{L} = -\frac{1}{4} \left ( \partial \wedge \B + \frac{1}{n \cdot \partial} \overline{n \wedge J} \right )^2 + \frac{1}{2} \tilde m^2 \B_\mu \B^\mu \, .
\eea
As is the case when this formalism was first introduced by Dirac, the electromagnetic field strength is modified to be
\bea \label{Eq: field}
F = - \overline{\partial \wedge \B} + \frac{1}{n \cdot \partial} n \wedge J .
\eea
When there is an electron present, the solution for $\overline{\partial \wedge \B}$ has a singularity that describes the infinitely thin solenoid leaving the electron.  By considering the exact combination shown in Eq.~\ref{Eq: E field}, this singularity in $\overline{\partial \wedge \B}$ cancels the singularity in $\frac{1}{n \cdot \partial} n \wedge J$, leaving just the electric field of the electron without any Dirac string attached to it.

We expect that Eq.~\ref{Eq: two} is more general than Eq.~\ref{Eq: one} and Eq.~\ref{Eq: three}, as it can easily be extended to accommodate non-straight Dirac strings of different particles pointing in different directions~\cite{Ripka:2003vv}.  Unfortunately, we cannot prove that these extensions remain local, though we suspect that they are.

\paragraph{Two field formalism}

At the price of introducing an extra gauge field $\B_\mu$, everything can be made to be completely local.  This two field formalism was originally introduced by Zwanziger~\cite{Zwanziger:1970hk}.
\bea \label{Eq: three}
\mathcal{L} &= -\frac{1}{2 n^2} \left ( n \cdot (\partial \wedge A) \right )^2  
-\frac{1}{2 n^2} \left ( n \cdot (\partial \wedge A) \right ) \cdot \left ( n \cdot (\overline{\partial \wedge \B}) \right )  \\
& - \frac{1}{2 n^2} \left ( n \cdot (\partial \wedge \B) \right )^2
 +\frac{1}{2 n^2} \left ( n \cdot (\partial \wedge \B) \right ) \cdot \left ( n \cdot (\overline{\partial \wedge A}) \right )  \\
&  + A_\mu J^\mu + \frac{1}{2} \tilde m^2 \B_\mu \B^\mu \, .
\eea
The price of this two field formalism is that the usual electric and magnetic fields are described by
\bea \label{Eq: zwag}
F = \frac{1}{n^2} \left ( n \wedge (n \cdot (\partial \wedge A) ) - \overline {n \wedge (n \cdot (\partial \wedge \B) )} \right ).
\eea
In words, Eq.~\ref{Eq: zwag} means that $A_\mu$ describes $\vec E \parallel n$ and $\vec B \perp n$ while $\B_\mu$ describes $\vec B \parallel n$ and $\vec E \perp n$.  While this formalism is a bit clunky, it does demonstrate that the theory is completely local.

Eq.~\ref{Eq: one}, Eq.~\ref{Eq: two}, and Eq.~\ref{Eq: three} are connected.  Starting with Eq.~\ref{Eq: three}, if we integrate out $\B_\mu$, we obtain Eq.~\ref{Eq: one}.  If we instead integrate out $A_\mu$, we arrive at Eq.~\ref{Eq: two}. We will mainly make use of Eq.~\ref{Eq: two}, which is the easiest to extend to a large number of charges.

\section{Phenomenology} \label{Sec: pheno}

In this section, we discuss the main phenomenological consequences of the Lagrangian in Eq.~\ref{Eq: two}. The solution to the modified Maxwell's equations in the presence of a current $J$ is
\bea \label{Eq: sol}
\B^\mu = \frac{\epsilon^{\mu \nu \rho \sigma}}{\left ( \Box + \tilde m^2 \right ) \left ( n \cdot \partial \right ) } \partial_\nu n_\rho J_\sigma .
\eea
In the following, we consider the case where there is a Dirac string stretched throughout all of space and demonstrate that the unphysical Dirac string acquires a physical flux tube attached to it. We also calculate the potential energy between an electron and positron both at rest. Then, we demonstrate that the magnetic field is screened on distances larger than $1/\tilde m$ and that the Bianchi identity is violated.

\subsection{Electric field}

\paragraph{Dirac string} Electrostatic solutions of Maxwell's equations can be found in the presence of a Dirac string.  We find that while the Dirac string is unphysical, it acquires a physical flux tube surrounding it.

To get an infinite Dirac string pointing along the z-axis, we will choose the source $J^0 = e \, \delta^3(\vec r + \frac{R}{2} \hat z) - e \, \delta^3(\vec r - \frac{R}{2} \hat z )$, take $n^\mu$ to point in the z direction so that the Dirac string stretches between the two charged particles, and take the limit where $R \rightarrow \infty$.  With this, and using Eq.~\ref{Eq: sol} and Eq.~\ref{Eq: field}, we find that
\bea \label{Eq: E field}
\vec B = 0 \qquad \vec E = \frac{e \tilde m^2}{2 \pi} K_0 (\tilde m r) \, \hat z \, ,
\eea
where $K_0$ is the modified Bessel function and $r$ is the cylindrical distance from the Dirac string.

As claimed, we see that the delta function electric field of the Dirac string has vanished, indicating that it is indeed unphysical.  More surprisingly, we see that the unphysical Dirac string has acquired a physical string counterpart.  Eq.~\ref{Eq: E field} is the electromagnetic field of a standard Abrikosov vortex solution in the limit that the Higgs mode mass is taken to infinity~\cite{Abrikosov:1956sx} (see also~\cite{tinkham2004introduction,East:2022rsi}).  We see that there is a Dirac string at the center of every confined flux tube.  The flux contained in the string is equal and opposite that of the Dirac string.

\paragraph{Potential energy of an electron-positron pair} Taking Eq.~\ref{Eq: two}, completing the square and integrating out the dual photon, we find
\bea
\mathcal{L} = J_\mu \left ( -\frac{1}{2} \frac{g^{\mu \nu}}{\Box + \tilde m^2} - \frac{1}{2} \frac{n^2}{(n \cdot \partial)^2} \left ( \frac{m^2}{\Box + m^2} \right ) \left ( g^{\mu \nu} - \frac{n^\mu n^\nu}{n^2} \right )\right ) J_\nu \, . \nn
\eea
We will choose the source to be an electron positron pair $J^0 = e \, \delta^3(\vec r) - e \, \delta^3(\vec r + R \hat z )$ and take $n^\mu$ to point in the z direction.  Plugging this in, we find that the energy of this configuration is
\bea
V(R) &= - \frac{e^2}{(2 \pi)^3} \int d^3 k \, e^{-i \vec k \cdot \vec z R} \left ( \frac{1}{k^2 +\tilde m^2} + \frac{\tilde m^2}{k^2 + \tilde m^2} \frac{1}{(\hat z \cdot \vec k )^2} \right ) \nn\\
&= - \frac{e^2}{4 \pi R} e^{-\tilde m R} + \frac{e^2 \tilde m^2 R}{4 \pi^2} \int_0^{\Lambda R} dy \frac{1}{y^2 + \tilde m^2 R^2} (\cos y + y \, \text{Si} ( y) ) 
\eea
where $\text{Si}$ is the sine integral and the momentum integral is cut off at a UV scale $\Lambda$. In the $R \gg 1/\tilde{m}$ limit, the potential is 
\begin{equation}\label{eq:longdistance}
 V(R) \approx - \frac{e^2}{4 \pi R} e^{-\tilde m R} + \frac{e^2 \tilde m^2 R}{4 \pi} \log \frac{\Lambda}{\tilde m}.
\end{equation}
The logarithmically divergent tension is the standard log divergent tension found in all strings.  The lack of a counter-term to absorb this log divergence indicates that our Lagrangian is missing a degree of freedom, namely the dynamical flux tubes.
Given that the $U(1)$ theory has a Landau pole and that the Planck scale provides another possible cutoff scale, this logarithm is not particularly large. At small $R \ll 1/\tilde m$, the potential is
\begin{equation}\label{eq:shortdistance}
 V(R) \approx - \frac{e^2}{4 \pi R}  + \frac{e^2 \tilde m^2 R}{4 \pi} \log \Lambda R,
\end{equation}
where the same logarithmic dependence on the UV scale appears. The physical interpretation of this potential is pretty clear.  At short distances, the electron-positron pair behaves as expected from Maxwell's equations, with corrections from the existence of an infinitely thin string core.  At long distances, the electric flux confines into a flux tube that stretches between the two particles and has string tension $\mu = \frac{e^2 \tilde m^2}{4 \pi} \log \frac{\Lambda}{\tilde m}$ providing a constant force between the two particles.

\subsection{Magnetic field} \label{Sec: screen}

\paragraph{Violation of the Bianchi identity} Given Eq.~\ref{Eq: sol}, a time-independent space-like current $J_i$ produces a non-zero $\B^0$, which leads to a modification of the Bianchi identity $\vec \nabla \cdot \vec B = 0$ to 
\begin{equation}\label{eq:bianchi}
\vec \nabla \cdot \vec B = \tilde{m}^2 \B^0,
\end{equation}
suggesting that magnetic field lines are no longer closed in the presence of a persistent current. This suggests that a dipolar magnetic field generated from, for example, a current loop, can be screened at long distances. 

\paragraph{Screening of a magnetic field} Consider a current $I$ going in the z-direction and solve for the magnetic field as a function of distance from the current using Eq.~\ref{Eq: sol} and Eq.~\ref{Eq: field}.  We find that the electromagnetic field surrounding a current is
\bea
\vec E = 0 \qquad \vec B = \frac{I}{2 \pi r}  \, \left ( \tilde m r \, K_1(\tilde m r) \right ) \, \hat \phi .
\eea
At close distances, $\tilde m r \ll 1$, the new factor $ \left ( \tilde m r \, K_1(\tilde m r) \right ) \rightarrow 1$ and we get the usual result for a magnetic field around a wire.  However, at large distances, $\tilde m r \gg 1$, an exponential suppression kicks in $ \left ( \tilde m r \, K_1(\tilde m r) \right ) \rightarrow e^{-\tilde m r} \sqrt{\pi \tilde m r/2} $.  Thus we see that magnetic fields generated by currents are screened in the long-distance limit.  In the limit of large $\tilde{m}$ it is tricky to treat a densely packed flow of charged particles in a wire with a complicated shape such as a toroid. However, given the strength of the astrophysical constraint (see section~\ref{sec:astro} for more details), such a detailed computation is likely unnecessary for any realistic system in the laboratory.

\paragraph{Lack of screening of a magnetic field} Just as a Proca mass screens most but not all electric fields, a dual photon mass screens most but not all magnetic fields. There is a longitudinal mode. For example, in the absence of a source, Eq.~\ref{Eq: two} has a propagating plane wave solution
\bea \label{eq:longitudinal}
B^i = a \cos \left ( \omega t - \vec k \cdot \vec r \right ) \qquad \omega^2 = k^2 + \tilde m^2 .
\eea
This propagating plane wave has a non-zero $\vec B$ field and a velocity suppressed $\vec E$ field.  From this, we see that it is possible to have a $\vec B$ field coherent on a distance $ \sim 1/k \gg 1/\tilde m$ at the price of having time variation on time scales $\sim 1/\tilde m$. 

To conclude, whereas the violation of $\vec \nabla \cdot \vec B = 0$ has a direct phenomenological consequence, the screening of long-range magnetic fields or the lack of, though very closely related, does not always offer a conclusive test. 

\section{Experimental constraints} \label{Sec: constraint}

There are three main ways to look for a dual photon mass.  There are the dispersion relationship of the photon, modifications to the $1/r^2$ force, and the screening of magnetic fields.  In this section, we will describe constraints resulting from searches of these kind.  Similar to the case of a normal Stuckelberg mass, the strongest bounds come from astrophysics (for a review see Refs.~\cite{Goldhaber:1971mr,Goldhaber:2008xy}).  However, these astrophysical bounds are fraught with order of magnitude or more uncertainty and thus we will consider separately the more concrete laboratory constraints and the astrophysical bounds.  

\subsection{Laboratory Constraints}

There are laboratory-based constraints on the dispersion relationship of light and violations of the $1/r$ Coulomb potential shown in Eq.~\ref{eq:shortdistance}, whereas a precision experiment that tests a fundamental violation of the Bianchi identity shown in Eq.~\ref{eq:bianchi} has likely not been done.

\paragraph{Schumann resonance}
The simplest constraint on the dual photon mass comes from propagation of light.  
Schumann resonances are atmospheric electromagnetic waves trapped between the Earth's surface and the ionosphere.  Kroll's measurement of very low frequency waves places a constraint on the dual photon mass~\cite{Kroll:1971wi,Malta:2022zdb}
\begin{equation}
 \tilde{m} \lesssim 3\times 10^{-13} \,{\rm eV}.
\end{equation}

\paragraph{Lamb shift}  In the limit of $r\ll 1/{\tilde{m}} $, the correction to the Lamb shift is approximately
\begin{equation}
\Delta E \sim \frac{e^2 \tilde{m}^2 }{4\pi} a_0 \log[\Lambda a_0] \, ,
\end{equation}
where $a_0$ is the Bohr radius. Given the precision of the recent measurement~\cite{Bezginov:2019mdi}, we find a limit on the magnetic mass of $\tilde{m} \lesssim \,{\rm meV}$. A similar limit can be obtained from similar measurements with muons~\cite{Mu-MASS:2021uou}, and we expect precision measurements of atomic energy levels to give similar constraints. It is clear that these measurements are not competitive against long-range magnetic field measurements. 

Precision tests of Coulomb's Law, see e.g. Ref.~\cite{Williams:1971ms}, do not apply easily to the case of a dual photon mass.  Traditionally, tests of Coulomb's Law are Cavendish-type experiments that utilize $\int da \cdot \vec E = 0$ to show that the electric field is zero inside of a uniformly charged sphere.  Unfortunately, a dual photon mass preserves Gauss's law, and thus these experiments do not apply in a straightforward manner. These Cavendish-type experiments~\cite{Williams:1971ms} {\it do not} test violations of the inverse-square law but instead test violations of Gauss's law.

\paragraph{Magnetic (Aharonov-Bohm) experiment} A standard lab-based Aharonov-Bohm experiment tests magnetic fields at distances $\sim$ meter giving a bound $\tilde{m} \lesssim \, 1/{\rm meter} \sim 10^{-7} \, {\rm eV}$. An experiment similar to the one proposed in Ref.~\cite{Boulware:1989up}, including several designed to search for axion dark matter~\cite{Kahn:2016aff}, can be modified to extend this limit to $\tilde{m} \lesssim  10^{-12} \, {\rm eV}$. We leave a more detailed study to future work.

\subsection{Astrophysical constraints}\label{sec:astro}

\paragraph{Screening stellar magnetic fields} The presence of a long-range unscreened dipole magnetic field of the Earth (Jupiter) places a constraint on the photon mass of order $\tilde{m} < 1/R_{\rm earth} \approx 10^{-14} \,{\rm eV}$ ($\tilde{m} < 1/R_{\rm Jupiter} \approx 10^{-15} \,{\rm eV}$).   This is the strongest known constraint on the mass of the photon that explicitly tests $\vec \nabla \cdot \vec B \ne 0$.  Given the abundance of data on the magnetic fields of these planets, one can likely use the data to evaluate $\int \vec da \cdot \vec B$ around the entire planet as an explicit test of $\vec \nabla \cdot \vec B \ne 0$. A more detailed analysis can likely significantly improve this limit. 

\paragraph{Screening galactic magnetic field and a new experiment} The strongest constraint on the usual photon mass comes from the existence of galactic magnetic fields with a magnitude $\mu G$ that are coherent over distances of $\sim$ kpc~\cite{Chibisov:1976mm,Adelberger:2003qx} that extend off of the galactic plane.  If this magnetic field is generated by currents in the galactic disk, then we can use the results of Sec.~\ref{Sec: screen} to impose a bound.  The fact that the galactic magnetic field has not been exponentially damped implies that
\bea
\tilde m \lesssim \frac{1}{\rm kpc} \sim 10^{-26} \, \text{eV} . 
\eea

A loophole in the previous bound is that the galactic magnetic field can also receive contributions from very non-relativistic photons. In this case, the resulting magnetic field would oscillate as a function of time instead of being screened.  While this possibility is exotic and implausible, it is not excluded.  We are not aware of a study that directly limits the violation of $\vec \nabla \cdot \vec B = 0$ on galactic scale.  Because of this, we propose closing this loophole by studying the time dependence of the Faraday rotation resulting from an oscillating longitudinal mode shown in Eq.~\ref{eq:longitudinal}. An experiment along these lines would be to measure the direction and magnitude of the galactic magnetic field and measure it again a year later to confirm that the magnitude and direction have not changed.  If the time dependence can be demonstrated to be longer than a month, the resulting constraint would likely be stronger than the other constraints we consider.
\newline

\section{Conclusion} \label{Sec: conclusion}

In this article, we discussed a new IR phase for QED where electric fields are confined and magnetic fields are screened.  This phase occurs when a mass term is added for the dual photon.  After writing down the Lagrangian, we discussed the physical consequences of this term and placed an experimental constraint on $\tilde m$.  The physical consequence is that the electric field of charged particles is confined into a flux tube that connects it to an oppositely charged particle.  Meanwhile, magnetic fields are screened and acquire an exponential suppression $e^{- \tilde m r}$.  The best experimental constraints are astrophysical in nature, but are extremely uncertain and contain possible loopholes.  We propose a simple experiment to improve upon these bounds.
It would be interesting if a more complete second quantized, local and Lorentz invariant Lagrangian which describes both electrically and magnetically charged particles could be constructed. In that case, we can extend our discussions beyond steady states that are semi-classical.

\section*{Acknowledgements}

We thank Asimina Arvanitaki, Savas Dimopoulos, Gustavo Marques-Tavares, Riccardo Rattazzi and Raman Sundrum for useful conversations. AH is supported by the NSF grant PHY-1914480 and by the Maryland Center for Fundamental Physics (MCFP). Research at Perimeter Institute is supported in part by the Government of
Canada through the Department of Innovation, Science and Economic Development Canada and by
the Province of Ontario through the Ministry of Colleges and Universities.

\appendix

\bibliographystyle{apsrev4-1}
\bibliography{Refs}

\end{document}